\documentclass[preprint,12pt]{elsarticle}
\pdfoutput=1

\usepackage{amssymb}
\usepackage{amsmath}
\usepackage{graphicx}
\usepackage{enumerate}
\usepackage{url} 
\usepackage{hyperref}
\usepackage[figuresright]{rotating}
\usepackage{graphicx}
\usepackage{enumerate}
\usepackage{natbib}
\usepackage{url} 
\usepackage{hyperref}
\usepackage[figuresright]{rotating}
\usepackage{color}
\usepackage{dsfont}
\usepackage{amsthm}
\usepackage{multirow}
\usepackage{caption}
\usepackage{float}
\usepackage{subcaption}
\usepackage{threeparttable}

\RequirePackage{amsthm,amsfonts,amssymb}

\journal{arXiv}

\begin{document}

\begin{frontmatter}

\title{SleepEGAN: A GAN-enhanced Ensemble Deep Learning Model for Imbalanced Classification of Sleep Stages}

\author[label1,label2]{Xuewei Cheng}
\ead{xwcheng@csu.edu.cn}

\author[label3]{Ke Huang}
\ead{khuan049@ucr.edu}

\author[label1]{Yi Zou}
\ead{zy6868@csu.edu.cn}

\author[label3]{Shujie Ma\corref{mycorrespondingauthor}}
\cortext[mycorrespondingauthor]{Corresponding author}
\ead{shujie.ma@ucr.edu}

\address[label1]{School of Mathematics and Statistics,
Central South University, Changsha, China.}

\address[label2]{School of Mathematics and Statistics,
Hunan Normal University, Changsha, China.}

\address[label3]{Department of Statistics, University of California, Riverside, U.S.A.}


\begin{abstract}
Deep neural networks have played an important role in automatic sleep stage classification because of their strong representation and in-model feature transformation abilities. However,  class imbalance and individual heterogeneity which typically exist in raw EEG signals of sleep data can significantly affect the classification performance of any machine learning algorithms. To solve these two problems, this paper develops a generative adversarial network (GAN)-powered ensemble deep learning model, named SleepEGAN, for the imbalanced classification of sleep stages. To alleviate class imbalance, we propose a new GAN (called EGAN) architecture adapted to the features of EEG signals for data augmentation. The generated samples for the minority classes are used in the training process. In addition, we design a cost-free ensemble learning strategy to reduce the model estimation variance caused by the heterogeneity between the validation and test sets, so as to enhance the accuracy and robustness of prediction performance. We show that the proposed method can improve classification accuracy compared to several existing state-of-the-art methods using three public sleep datasets.
\end{abstract}


\end{frontmatter}

\section{Introduction}\label{S1}

Sleep plays a vital role in mental and physical well-being throughout an individual's life \cite{estrada2006itakura, aboalayon2016sleep}. According to the research in \cite{jahrami2021sleep,estrada2009eeg,estrada2010eeg} and the American Sleep Association, about 35.7\% of people in the world and 50-70 million adults in the United States have a sleep disorder. The lack of sleep can cause negative cognitive, emotional, and physical effects \cite{liu2022extracting}. In recent years, sleep stage classification has gained wide attention in the machine learning community \cite{aboalayon2016sleep,boostani2017}, as it is crucial for understanding the quality and quantity of sleep and for diagnosing and treating various sleep disorders \cite{2020Sleep,2023Quantifying}.

Sleep stage scoring is generally performed based on polysomnogram (PSG), which is considered the gold standard for evaluating human sleep \cite{fraiwan2012automated}. PSG monitors many body functions during sleep, including brain activity (electroencephalogram, EEG), eye movements (electrooculogram, EOG), muscle activity (electromyogram, EMG), and heart rhythm (electrocardiogram, ECG). Single-channel EEG signals have been popularly used for sleep stage scoring because they are convenient and less expensive to be monitored and collected \cite{ks2017}. Specifically, EEG recordings are typically segmented into epochs of 30 seconds,  and each epoch is manually labeled by sleep specialists and then classified into one of five stages: Wake (W), Rapid eye movement (REM), and three non-REM stages (N1, N2, N3), following the AASM (American Academy of Sleep Medicine) guidance \cite{berry2012aasm}. The task of the manual classification process is labor-intensive and prone to experts' subjective perception \cite{ronzhina2012sleep}. To this end, an automatic classification system for sleep stages can alleviate these problems and assist sleep specialists \citep{2022Automatic}.  
In recent years, the deep convolutional neural networks (CNNs) together with recurrent neural networks (RNNs) or long short-term memory (LSTM) networks \cite{supratak2017deepsleepnet,mousavi2019sleepeegnet,supratak2020tinysleepnet,eldele2021attention,fiorillo2021deepsleepnet,huang2022improved,zhao2022sleepcontextnet,phyo2022transsleep} have been successfully applied to sleep stage classification, as they can effectively learn frequency and time domain signals \cite{hassan2016decision,fraiwan2012automated} from raw EEG epochs. 

However, the {\it class imbalance} and {\it individual heterogeneity} of EEG signals, which are two common problems in sleep data,  have not been well-addressed in the literature. To be specific, the sleep duration at all stages is not evenly distributed. Stage N2 generally occupies most of the sleep time (40.3\%) and stage N1 only accounts for 6.3\% in the sleep-EDF-v1 dataset \cite{goldberger2000physiobank}. In general, the imbalance of data can seriously affect the classification performance \cite{liu2022extracting}. The heterogeneity is another challenge emerging from the raw data when they originate from different examining environments \cite{aboalayon2016sleep}, channel layouts, or recording setups \cite{phan2022automatic}. For example, patients often stay overnight in a sleep laboratory with adhesive electrodes and wires attached to their heads to measure sleep patterns, and this uncomfortable environment can affect their sleep quality \cite{lan2015using}, causing the heterogeneity of EEG signals. As a result, we may not have a good generalization ability on the test set based on the model parameters selected by the validation set.

To solve the aforementioned problems, this paper develops a generative adversarial network (GAN)-enhanced ensemble deep learning model, named SleepEGAN, for the imbalanced classification of sleep stages. To alleviate class imbalance, we propose a new GAN (called EGAN) architecture adapted to the features of EEG signals for data augmentation. The generator and discriminator models in our GAN are motivated by the deep neural networks called TinysleepNet proposed in \cite{supratak2020tinysleepnet}. TinysleepNet is originally used for classifying sleep stages and is shown to have a great capability of extracting features from raw EEG signals and learning their temporal transition rules using only four convolutional layers and a single LSTM layer. It achieves a good balance between generalization and parsimony while preserving its ability to learn the structure of EEG signals. We take advantage of its model structure and design a modified version of the TinysleepNet model used for the generator and discriminator in our EGAN architecture, specially tailored for the purpose of EEG signal augmentation.  

Our proposed EGAN model is shown to be an effective and efficient tool to generate EEG signals for small classes of sleep stages such as for stage N1 to match the number of samples in the large classes. It is worth noting that in the literature, a few works \cite{zhou2022alleviating,hartmann2018eeg} directly employ the existing GAN methods to generate EEG signals, such as the naive GAN and the Wasserstein GAN \cite{arjovsky2017wasserstein} originally proposed for image generation, which may heavily rely on the convolutional layers and thus possibly neglect the temporal and transitional features of EEG signals. Next, we design a new classification network structure SleepEGAN to classify the sleep stages with the augmented data.  The generated
samples for the minority classes are used in the training process of classification. Our SleepEGAN combines the advantages of the VGG16 \cite{simonyan2014very} and Tinysleepnet \cite{supratak2020tinysleepnet} models: the former enhances the network's signals via increasing the number of filters, while the latter learns temporal features by adding LSTM layers. Both of them are small, efficient, and computationally convenient classification networks. 

To tackle the problem of individual heterogeneity of EEG signals, we design a cost-free ensemble algorithm. Ensemble learning is a proven favorable and effective strategy to handle heterogeneous data \cite{jin2004switch}. It uses multiple diverse classifiers to achieve better generalization performance than a single learner to reduce prediction variance \cite{zhou2017deep}. Throughout the training process, we retain the model parameters obtained in the epochs from the top 10 models chosen based on the prediction accuracy on the validation set instead of keeping only one set of model parameters having the best prediction. The stage prediction on the test set is based on the ensemble result of these 10 models, as the model parameters in different updated epochs are heterogeneous, and they can perform well in the validation set, satisfying two sufficient conditions for a nice ensemble: accurate and diverse \cite{zhou2012ensemble}. It is worth noting that we only save the model parameters in each epoch during the training process, and then build an ensemble model using the retained parameters from the chosen models evaluated in the validation set. As a result, our ensemble learning procedure does not increase any training costs, as it does not require training any additional models compared to the conventional deep neural network algorithms without ensemble learning.

The rest of the paper is organized as follows. Section \ref{S2} introduces the proposed GAN-enhanced ensemble deep learning model. The proposed method is illustrated on three real sleep datasets with the numerical results reported in Section \ref{S3}. Concluding remarks are given in Section \ref{S4}.

\section{Propose method}\label{S2}
In this section, we introduce the proposed GAN-based ensemble deep learning model (SleepEGAN) for the imbalanced classification of sleep stages. Our method contains three steps:
\begin{itemize}
    \item we design a new GAN (EGAN) to generate samples for the minority classes so that the sample size of each class is balanced on the training set;
    \item we elaborately build a classification network architecture based on convolutional and LSTM layers;
    \item we develop an ensemble learning strategy without  additional computational cost to reduce the variance of the model prediction caused by heterogeneity.
\end{itemize}

\subsection{Data Augmentation with EGAN}
Signal and sequence augmentation is considered as a primary technique to synthesize new training data from the original data for each training epoch \cite{supratak2020tinysleepnet} to alleviate the class imbalance problem. We first use this technique to synthesize new signal patterns for each training epoch by signal augmentation and generate new batches of multiple sequences of EEG epochs in the mini-batch gradient descent by sequence augmentation. In addition, the weighted cross-entropy loss function is also introduced to mitigate the class imbalanced problem by setting the weight for the N1 stage to 1.5 and others to 1. However, these strategies of data augmentation cannot completely solve the imbalanced problem, and the learned deep model still prioritizes the majority class.

To solve the above problem, we propose a new generative adversarial network (EGAN) to learn the probability distribution for generating raw EEG epochs, as an advanced strategy for data augmentation. The generator of GAN is able to generate more samples from the estimated probability distribution. Different from parametric and nonparametric density estimators, where the density function is explicitly defined in a relatively low dimension, GAN can be viewed as an implicit density estimator in a much higher dimension \cite{goodfellow2020generative}. 

According to the invariance structure of data, data augmentation by GAN implicitly enlarges the training dataset by sampling original data and generating new data, which usually regularizes the model effectively \cite{fan2020statistical}. However, the naive GAN and other extensions \cite{radford2015unsupervised,arjovsky2017wasserstein}, which are originally designed for image generation, may not work well for EEG signals as they do not consider the temporal and transitional features of EEG data. Therefore, we design a new GAN architecture (EGAN) tailored for EEG signal generation. The proposed EGAN has two features: it is able to extract the representative temporal components from the high-dimensional features, and then automatically learn the transition rules of the EEG signals.

\begin{figure*}[htbp]
    \centering
    \includegraphics[scale=0.4]{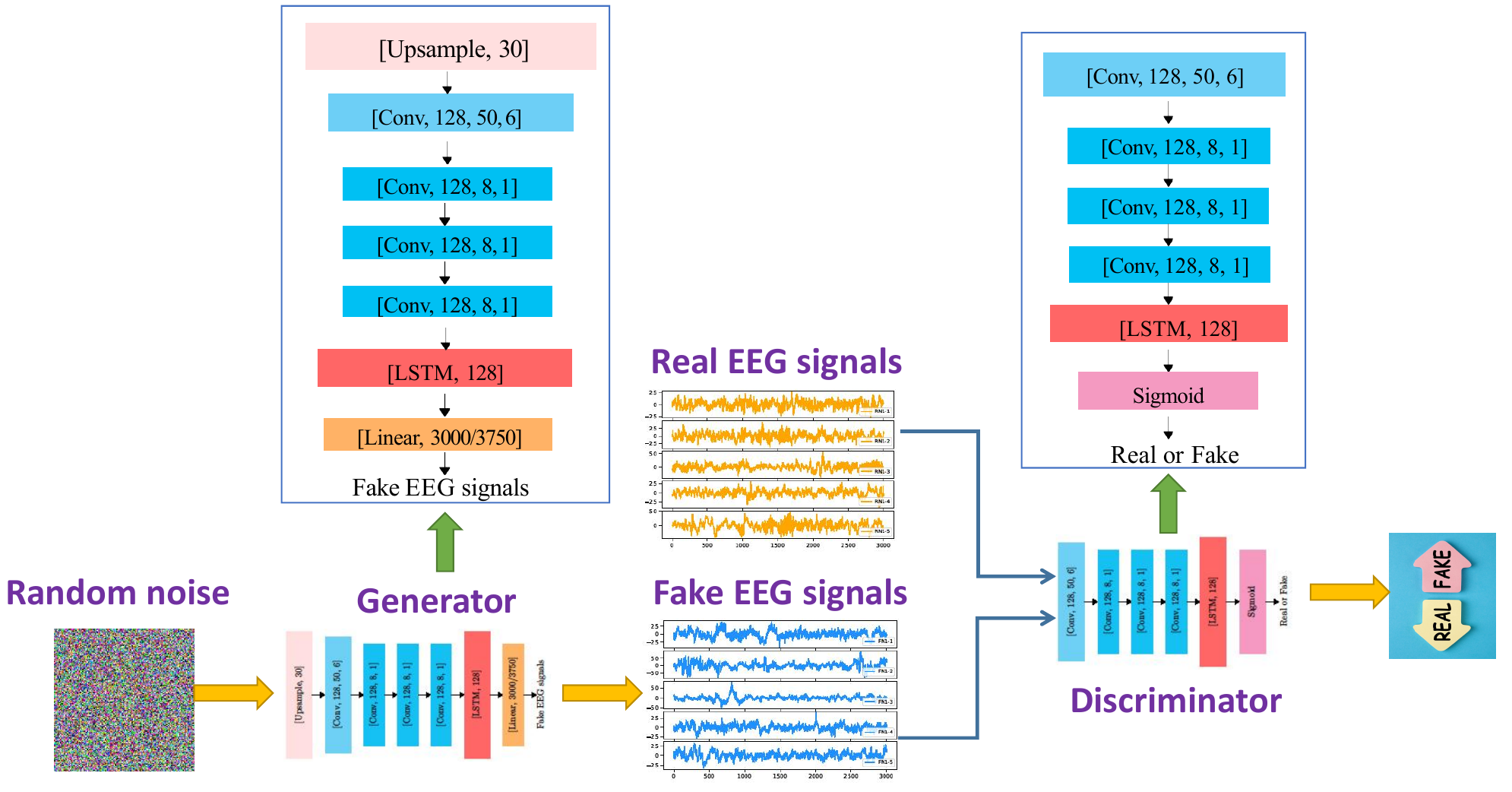}
    \caption{The main structure of EGAN. For simplicity, the pooling and dropout layers are ignored here.}\label{F5}
\end{figure*}

Inspired by Tinysleepnet originally proposed for sleep stage classification, we design a 4+1 learning framework, where the first four convolutional layers are used to extract frequency signals, followed by an LSTM layer to learn temporal information. The main structure of EGAN is presented in Figure \ref{F5}. However, we need to modify some details to make the network possible to achieve the high-quality task of sample generation. Specifically, we make the following contributions:

\begin{itemize}

\item for the generator, we add an Upsample layer to expand the 100-dimensional (or 125-dimensional) noise to 3000 (or 3750) dimensions as inputs;
\item for the generator, after the LSTM layer processing, we add a fully connected layer so that it transforms the learned features into 3000-dimensional (or 3750-dimensional) EEG epochs with Tanh nonlinear activation functions;
\item for the discriminator, it only needs to discriminate between true and false, not its specific sleep stage, so we modify the activation function of the output layer from Softmax to Sigmoid;
\item we change the activation function of ReLU in Tinysleepnet to leaky ReLU in EGAN to make sure the gradient can flow through the entire architecture.
\end{itemize}

The proposed GAN model plays a vital role in balancing training samples among different classes, resulting in a superior prediction performance on the N1 stage in sleep data.

\subsection{Classification with SleepEGAN}

\begin{figure}[htbp]
    \centering
    \includegraphics[scale=0.65]{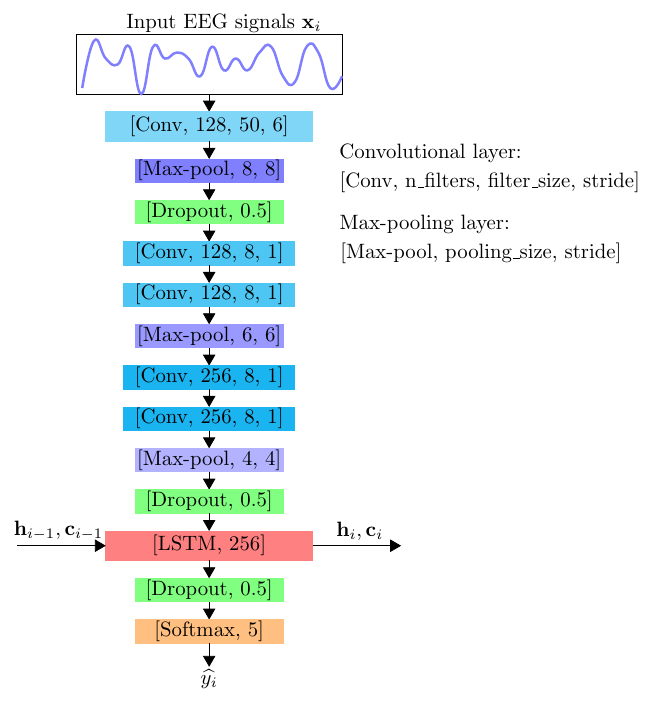}
    \caption{An overview classification network architecture of SleepEGAN. Each rectangular box represents one layer in the model, and the arrows indicate the flow of data from raw single-channel EEG epochs $\mathbf{x}_{i}$ to sleep stages $\widehat{y}_{i}$.}\label{F1}
\end{figure}

With the augmented data obtained from the previous step, next we design a new classification network SleepEGAN to process a sequence of single-channel EEG epochs and perform the classification of sleep stages (see Figure \ref{F1}). Our SleepEGAN combines the merits of VGG16 \cite{simonyan2014very} and Tinysleepnet \cite{supratak2020tinysleepnet}. Both VGG16 and Tinysleepnet are considered to be small and efficient classification networks. The former can enhance the network's signal by increasing the number of filters, while the latter is capable of learning the temporal patterns of EEG signals by using LSTM layers. Our SleepEGAN combines both advantages, using a 1+2+2 convolutional network to extract frequency features, followed by an LSTM layer to extract time domain features.

To be specific, we segment the EEG signals into $n$ epochs $\{\mathbf{x}_{1},...,\mathbf{x}_{n}\}$ of $E_{s}$ seconds, where $\mathbf{x}_{i} \in \mathds{R}^{E_{s} \times F_{s}}$ and $F_{s}$ is the sample rating for each second EEG. We obtain the predicted sleep stage $\widehat{y}_{i}$ in the test set using the epoch of $\mathbf{x}_{i}$ with the network parameters
trained using the training data set, where $\widehat{y}_{i} \in \{0,1,2,3,4\}$ corresponds to the five sleep stages W, N1, N2, N3 and REM, respectively.

The CNNs block $\text{CNN}_{\theta_{r}}$ that consists of five convolutional layers (1+2+2), interleaved with three max-pooling and two dropout layers, is firstly employed to learn time-invariant features from single-channel signals $\mathbf{x}_{i}$. Then, the LSTM layer $\text{LSTM}_{\theta_{s}}$ followed by a dropout layer is used to extract time-dependent information from processed features $\widetilde{\mathbf{x}}_{i}$ by CNNs block. The final out $y_{i}$ is activated by Softmax function $\sigma (\cdot)$ with parameter vector $\mathbf{v}$. In the procedure (\ref{E1}), $\theta_{r}$ and $\theta_{s}$ are the learnable parameters of the CNNs and LSTM, respectively, where $\mathbf{h}_{i}$ and $\mathbf{c}_{i}$ are output vectors of hidden and cell states of the LSTM layer after processing the features $\widetilde{\mathbf{x}}_{i}$.

\begin{equation}\label{E1}
    \begin{split}
        \widetilde{\mathbf{x}}_{i} &=\text{CNN}_{\theta_{r}}(\mathbf{x}_{i}),\\
        \mathbf{h}_{i},\mathbf{c}_{i} &=\text{LSTM}_{\theta_{s}}(\mathbf{h}_{i-1},\mathbf{c}_{i-1},\widetilde{\mathbf{x}}_{i}),\\
        y_{i} &= \sigma(\mathbf{v}\mathbf{h}_{i}).
    \end{split}
\end{equation}

We develop the SleepEGAN architecture with the goal of achieving a good balance between generalization and parsimony while preserving its ability to learn the structure of EEG signals. The strategies of bidirectional LSTM in Deepsleepnet \cite{supratak2017deepsleepnet}, 
multi-head attention in AttnSleep \cite{eldele2021attention},
dual-stream structure \cite{jia2021salientsleepnet} in SalientSleepNet and multi-scale extraction in \cite{liu2022extracting}
may enhance the representative ability of deep learning, but huge computational resources may incur. Our contribution is not to develop a deeper and more complicated neural network model with superior generalization ability. Instead, we would like to use limited resources to tackle the problems of individual heterogeneity and class imbalance.

\subsection{Ensemble learning}

For the classification of sleep stages, typically researchers use EEG signals from a number of individuals as the test set. Moreover, they split the individuals from the training data set into the training and validation sets. 
In this way, if there exists individual heterogeneity in the training and test sets, the model parameters selected by the validation set may not perform well on the test set. This heterogeneity can be inherent to the raw EEG data. To solve this problem, we introduce a cost-free ensemble learning strategy to improve the accuracy and stability of the prediction performance.

\begin{figure}[htbp]
    \centering
    \includegraphics[scale=0.75]{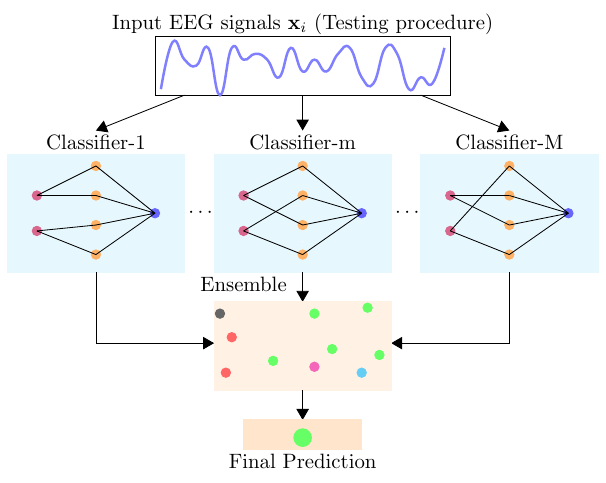}
    \caption{Ensemble illustration of predicted stage class. Different marks in the ensemble rectangle imply different predicted stage classes. The final stage prediction is obtained by classifier voting.}\label{F2}
\end{figure}

Ensemble learning is a machine learning paradigm where multiple learners are trained and combined for a specific task, achieving better generalization performance than single learners \cite{zhou2017deep}. However, the computational cost of ensemble learning can be much higher than that of a single classifier, especially when the ensemble is performed on deep neural networks. To address this issue, we develop a cost-free algorithm for ensemble learning. Specifically, we record the validation accuracy and F1-score for each training epoch, and we select the model parameters of the top $M$ classifiers ranked by their prediction performance evaluated on the test set.  Each model makes a separate prediction for each sample, and eventually, all classifiers take a majority vote on the final prediction class (see Figure \ref{F2}). By this method, we reduce the prediction variance and improve the stability of the prediction performance without paying any additional computational cost.

\section{Experimental results}\label{S3}

In this section, we first introduce the sleep datasets and experimental settings. Then, we show the classification performance of our proposed SleepEGAN, and compare it with several existing classification methods. 

\subsection{EEG Datasets}\label{S3-1}

We evaluate our method by three popular and public real-life sleep datasets, namely, Sleep-EDF-20, Sleep-EDF-78,
and SHHS (Sleep Heart Health Study) as shown in Table \ref{ta1}.

\begin{table*}[!ht]
    \centering
    \caption{The details of three datasets before and after data augmentation.}\label{ta1}
     \resizebox*{130mm}{!}{
    \begin{tabular}{ccccccccccc}
    \hline
        Datasets & Subjects & Channel & Sampling rate & Type & W & N1 & N2 & N3 & REM & Epochs \\ \hline
        \multirow{4}{*}{Sleep-EDF-20} & \multirow{4}{*}{20} & \multirow{4}{*}{Fpz-Cz} & \multirow{4}{*}{100Hz} & \multirow{2}{*}{Before} & 10197 & 2804 & 17799 & 5703 & 7717 & \multirow{2}{*}{44220} \\ 
        & ~ & ~ & ~ & ~ & ~  23.1\% & 6.3\% & 40.3\% & 12.9\% & 17.5\%  & ~\\ \cline{5-11}
        & ~ & ~ & ~  & \multirow{2}{*}{ After} & 10197 & \textbf{8120} & 17799 & 5703 & 7717 & \multirow{2}{*}{49536}   \\ 
        & ~ & ~ & ~ & ~ & 20.6\% & 16.4\% & 35.9\% & 11.5\% & 15.6\% & ~ \\ \hline
        \multirow{2}{*}{Sleep-EDF-78} & \multirow{2}{*}{78} & \multirow{2}{*}{Fpz-Cz} & \multirow{2}{*}{100Hz} & \multirow{2}{*}{Before} & 69824 & 21522 & 69132 & 13039 & 25835 & \multirow{2}{*}{199352} \\ 
        & ~ & ~ & ~ & ~ & 35.0\% & 10.8\% & 34.7\% & 6.5\% & 13.0\%  & ~ \\  \hline
        \multirow{4}{*}{SHHS} & \multirow{4}{*}{329} & \multirow{4}{*}{C4-A1} & \multirow{4}{*}{125Hz} & \multirow{2}{*}{Before} & 46319 & 10304 & 142125 & 60153 & 65953 & \multirow{2}{*}{324854} \\ 
        & ~ & ~ & ~ & ~ & 14.3\% & 3.2\% & 43.8\% & 18.5\% & 20.3\%  & ~ \\ \cline{5-11}
        & ~ & ~ & ~ & \multirow{2}{*}{After} & 46319 & \textbf{46272} & 142125 & 60153 & 65953 & \multirow{2}{*}{360822}  \\ 
        & ~ & ~ & ~ & ~ & 12.8\% & 12.8\% & 39.4\% & 16.7\% & 18.3\%  & ~ \\ \hline
    \end{tabular}
    }
\end{table*}

The dataset of Sleep-EDF \cite{goldberger2000physiobank} has two versions. One is Sleep-EDF-20 published in 2013 before the expansion, in which there were 39 PSG recordings from the study of age effects in healthy subjects (SC), collected from 20 subjects. The other one is Sleep-EDF-78, which expands the number of recordings from 39 to 153, including 78 subjects aged between 25-101 years (37 males and 41 females). These recordings are segmented into 30s epochs and manually labeled by sleep experts in light of R \& K (Rechtschaffen and Kales) manual \cite{
rechtschaffen1968manual}. We evaluated our model using the Fpz-Cz EEG channel provided in these PSG recordings with a sampling rate of 100 Hz.

We also use a larger sleep dataset named SHHS \cite{quan1997sleep,zhang2018national} which is a multi-center cohort study about cardiovascular and other sleep-disordered breathing diseases to evaluate the performance of the proposed method. 
The subjects of this dataset suffer from a wide range of diseases, such as lung diseases,  cardiovascular diseases and coronary diseases. 
Following the study of
\cite{fonseca2016cardiorespiratory,eldele2021attention}, we select 329 subjects with regular sleep from 6441 subjects for our experiments to reduce the effect of other diseases. In addition, we select the C4-A1 channel with a sampling rate of 125 Hz. 

\subsection{Experiment settings}\label{S3-2}

The 20-fold cross-validation (CV) scheme was employed to evaluate the prediction performance on the three datasets. In each fold, we further allocate 10\% of the training set into a validation set for evaluating the training model in case of over-fitting. The models that achieve the top $M$ overall accuracy are kept for evaluation with the test set. We use Adam optimizer with 200 epochs to train the classification model, where the learning rate, Adam's beta1, and beta2 are $10^{-4}$, 0.9 and 0.999, respectively. The mini-batch size is set as 8, 32, and 128 for Sleep-EDF-20, Sleep-EDF-78, and SHHS, respectively. The sequence length is 20. The number $M$ of learners is 10, which works well in the trade-off between diversity and accuracy.

For the generative networks (EGAN), we also use the Adam optimizer to train the generator and discriminator, where the learning rate, Adam's beta1, and beta2 are $2 \times 10^{-4}$, 0.5 and 0.999, respectively. 
For the generated tasks for Sleep-EDF-20 and SHHS,
we set up 660 and 843 training epochs. The batch sizes are 16 and 64, respectively. The input to the generator is a 100-dimensional (or 125-dimensional) noise for Sleep-EDF-20 (or SHHS).

We use three metrics to evaluate the performance of our proposed method, namely, overall accuracy (ACC), macro-averaged F1-score (MF1), and Cohen's Kappa coefficient ($\kappa$). The second one
is a common metric to evaluate the performance of imbalanced datasets, and the last one is used to evaluate the consistency of prediction results.

\subsection{Generation of minority class samples by EGAN}\label{S3-3}

\begin{figure*}[htbp]
	\centering
	\begin{minipage}{0.32\linewidth}
		\centering
		\includegraphics[scale=0.33]{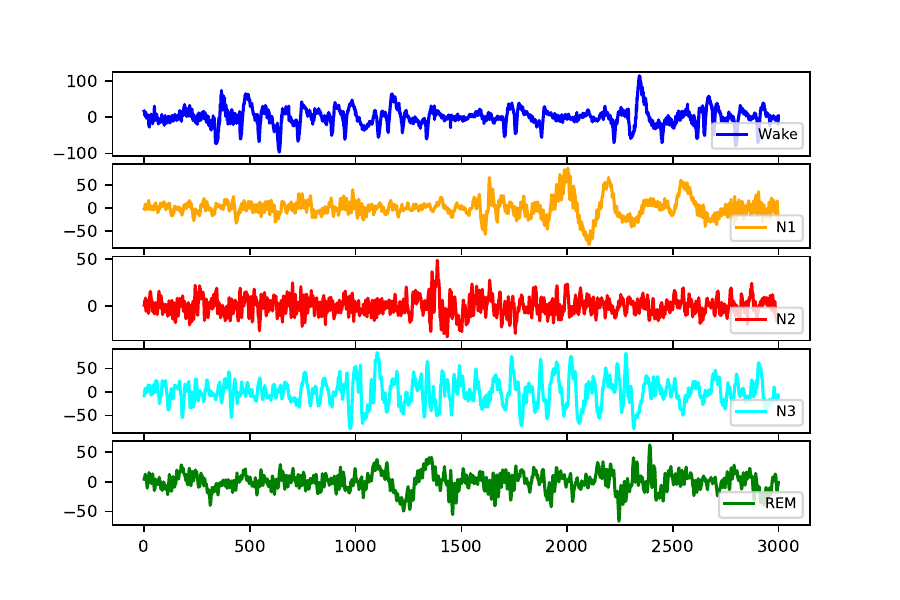}
		\caption{The real EEG signals
  in the five stages of Sleep-EDF-20.}
		\label{F3-1}
	\end{minipage}
	\begin{minipage}{0.32\linewidth}
		\centering
		\includegraphics[scale=0.33]{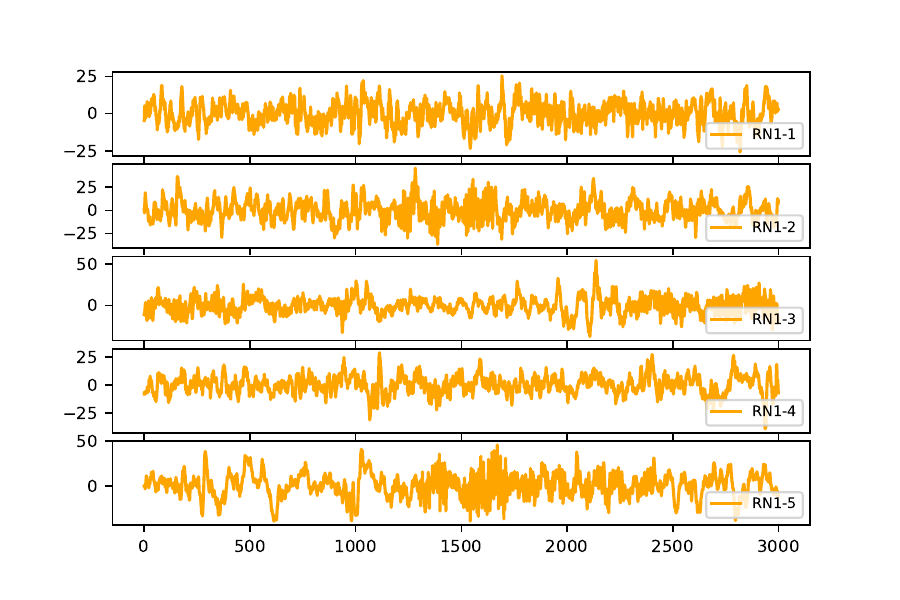}
		\caption{The real EEG signals
  in the N1 stage of Sleep-EDF-20.}
		\label{F3-2}
	\end{minipage}
 	\begin{minipage}{0.32\linewidth}
		\centering
		\includegraphics[scale=0.33]{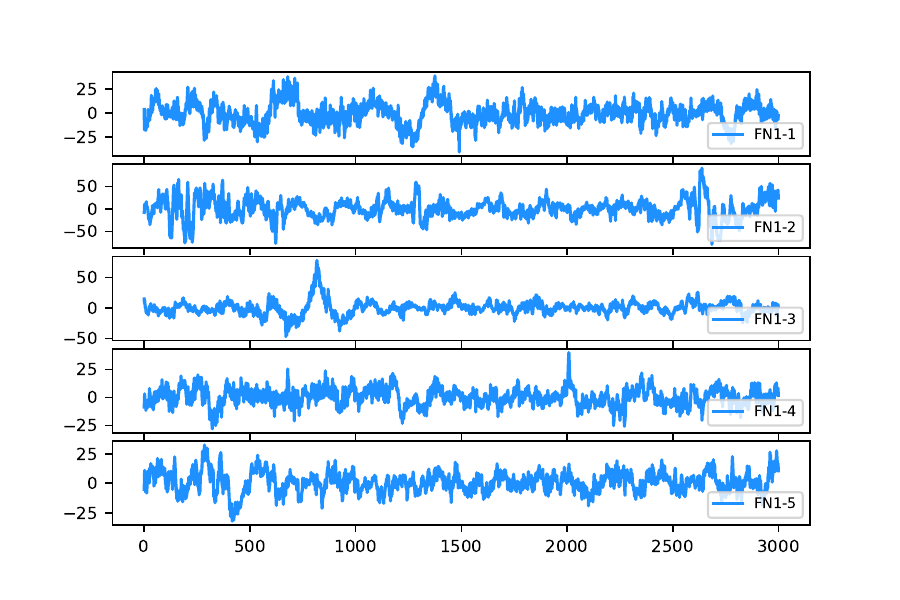}
		\caption{The generated fake EEG signals
  in the N1 stage of Sleep-EDF-20.}
		\label{F3-3}
	\end{minipage}
\end{figure*}

We generate EEG signals for the minority classes using our EGAN method for two sleep datasets Sleep-EDF-20 and SHHS, respectively, and the results of the generated samples are shown in Table \ref{ta1}. Specifically, for Sleep-EDF-20 and SHHS, N1 is the minority class, so we generate samples in the N1 stage to make the proportion of each class balanced. For Sleep-EDF-78, the sample size is balanced across sleep stages, so we drop out of the generation procedure. We generate samples for the smallest class to match the sample size of the penultimate class.

Figure \ref{F3-1} shows the EEG epochs in the five sleep stages for the dataset Sleep-EDF-20. Clearly, we can see that the EEG signals follow different patterns in the five sleep stages.  Moreover, Figures \ref{F3-2} and \ref{F3-3} show the real and the generated fake EEG signals in the N1 stage, respectively. We see that in general, the generated samples are quite similar to the real ones. The successful training of EGAN can make a good balance of different classes of sleep data for classification so as to improve prediction accuracy. It also indicates that EGAN has the ability to learn the distribution of temporal data, so it may have potential applications in signal denoising and detection. 

\subsection{Comparison of classification performance by different methods}\label{S3-4}

We evaluate the prediction performance of our SleepEGAN model against several state-of-the-art approaches. The comparison results among different methods are shown in Table \ref{ta2}. We observed that our sleepEGAN reasonably outperforms the other models in all three real datasets, thanks to the assistance of EGAN and ensemble learning.

\begin{table*}[!ht]
    \centering
    \caption{Comparison results between our method and other methods. The best performance on each dataset is highlighted in bold.}\label{ta2}
     \resizebox*{130mm}{!}{
    \begin{tabular}{ccccccccccc}
    \hline
        \multirow{2}{*}{Datasets} &
        \multirow{2}{*}{Methods}  & \multirow{2}{*}{Epochs} & \multicolumn{3}{c}{Overall Metrics} & \multicolumn{5}{c} {Per-class F1-Score (F1)}  \\ \cline{4-11}
        & & & Acc & MF1 & $\kappa$ & W & N1 & N2 & N3 & REM  \\ \hline
        \multirow{8}{*}{Sleep-EDF-20} & DeepSleepNet \cite{supratak2017deepsleepnet} & 41950 & 82.0 & 76.9 & 0.76 & 84.7 & 46.6 & 85.9 & 84.8 & 82.4 \\ 
        & IITNeT \cite{seo2020intra} & 42308 & 84.0 & 77.0 & 0.78 & 87.9 & 44.7 & 88.0 & 85.7 & 82.1 \\ 
        & SleepEEGNet \cite{mousavi2019sleepeegnet} & 42308 & 84.3 & 79.7 & 0.79 & 89.2 & 52.2 & 86.8 & 85.1 & 85.0 \\ 
        & ResnetLSTM \cite{sun2018deep} & 42308 & 82.5 & 73.7 & 0.76 & 86.5 & 28.4 & 87.7 & 89.8 & 76.2 \\ 
        & MultitaskCNN \cite{phan2018joint} & 42308 & 83.1 & 75.0 & 0.77 & 87.9 & 33.5 & 87.5 & 85.8 & 80.3 \\ 
        & AttnSleep \cite{eldele2021attention} & 42308 & 84.4 & 78.1 & 0.79 & 89.7 & 42.6 & 88.8 & \textbf{90.2} & 79.0 \\ 
        & Tinysleepnet \cite{supratak2020tinysleepnet} & 44220 & 85.4 & 80.5 & 0.80 & 90.1 & 51.4 & 88.5 & 88.3 & 84.3 \\ 
        & Our method & 44220 & \textbf{86.8} & \textbf{81.9} & \textbf{0.82} & \textbf{91.7} & \textbf{53.6} & \textbf{89.2} & 89.1 & \textbf{86.1} \\ \hline
        \multirow{7}{*}{Sleep-EDF-78} & DeepSleepNet \cite{supratak2017deepsleepnet} & 195479 & 77.8 & 71.8 & 0.70 & 90.9 & 45.0 & 79.2 & 72.7 & 71.1 \\ 
        & SleepEEGNet \cite{mousavi2019sleepeegnet} & 195479 & 74.2 & 69.6 & 0.66 & 89.8 & 42.1 & 75.2 & 70.4 & 70.6 \\ 
        & ResnetLSTM \cite{sun2018deep} & 195479 & 78.9 & 71.4 & 0.71 & 90.7 & 34.7 & 83.6 & 80.9 & 67.0 \\ 
        & MultitaskCNN \cite{phan2018joint} & 195479 & 79.6 & 72.8 & 0.72 & 90.9 & 39.7 & 83.2 & 76.6 & 73.5 \\ 
        & AttnSleep \cite{eldele2021attention} & 195479 & 81.3 & 75.1 & 0.74 & 92.0 & 42.0 & 85.0 & \textbf{82.1} & 74.2 \\ 
        & Tinysleepnet \cite{supratak2020tinysleepnet} & 199352 & 83.1 & 78.1 & 0.77 & 92.8 & 51.0 & 85.3 & 81.1 & 80.3 \\ 
        & Our method & 199352 & \textbf{83.8} & \textbf{78.7} & \textbf{0.82} & \textbf{93.1} & \textbf{51.7} & \textbf{85.8} & 81.2 & \textbf{82.0} \\ \hline 
        \multirow{6}{*}{SHHS} & DeepSleepNet \cite{supratak2017deepsleepnet} & 324854 & 81.0 & 73.9 & 0.73 & 85.4 & 40.5 & 82.5 & 79.3 & 81.9 \\ 
        & SleepEEGNet \cite{mousavi2019sleepeegnet} & 324854 & 73.9 & 68.4 & 0.65 & 81.3 & 34.4 & 73.4 & 75.9 & 77.0 \\ 
        & ResnetLSTM \cite{sun2018deep} & 324854 & 83.3 & 69.4 & 0.76 & 85.1 & 9.4 & 86.3 & 87.0 & 79.1 \\
        & MultitaskCNN \cite{phan2018joint} & 324854 & 81.4 & 71.2 & 0.74 & 82.2 & 25.7 & 83.9 & 83.3 & 81.1 \\ 
        & AttnSleep \cite{eldele2021attention} & 324854 & 84.2 & 75.3 & 0.78 & 86.7 & 33.2 & 87.1 & 87.1 & 82.1 \\ 
        & Our method & 324854 & \textbf{88.0} & \textbf{82.1} & \textbf{0.83} & \textbf{89.6} & \textbf{54.1} & \textbf{89.2} & \textbf{87.1} & \textbf{90.6}\\ \hline
    \end{tabular}
    }
\end{table*}

Specifically, the more imbalanced the dataset is, the better the performance of our method is after data augmentation. For example, for the dataset SHHS, after applying our EGAN and the ensemble learning, the size of N1 epochs increases from 10,304 to 46,272, and the F1-Score for the N1 class is improved from 40.5\% to 54.1\% by $13.6\%/40.5\%=33.6\%$ compared to the second best method.  The overall accuracy also improved to 88.0\%. In conclusion, the proposed SleepEGAN method has a promising performance for sleep stage prediction and is expected to work well for sleep data with imbalanced classes and individual heterogeneity.

\subsection{Ablation study}\label{S3-5}

Our method SleepEGAN contains two strategies to tackle the problems of class imbalance and individual heterogeneity.  To analyze the effectiveness of each strategy in our SleepEGAN, we provide an ablation study based on Sleep-EDF-20 as shown in Table \ref{ta3}. To be specific, we develop four model variants as follows.

\begin{itemize}
    \item Naive: only use main network structure to perform training process for classification. 
    \item Naive + EGAN: only use EGAN to generate naturalistic EEG epochs. 
    \item Naive + Ensemble: only use the ensemble strategy to enhance the prediction performance. 
    \item SleepEGAN: use both strategies to train EEG samples.
\end{itemize}

\begin{table}[!ht]
    \centering
    \caption{Ablation study conducted on Sleep-EDF-20 dataset.}\label{ta3}
    \resizebox*{110mm}{!}{
    \begin{tabular}{ccccccccc}
    \hline
        \multirow{2}{*}{Methods} & \multicolumn{3}{c}
          {Overall Metrics} & \multicolumn{5}{c} {Per-class F1-Score (F1)} \\ \cline{2-9}
        & Acc & MF1 & $\kappa$ & W & N1 & N2 & N3 & REM \\ \hline
        Naive & 85.1 & 79.8 & 0.80 & 90.1 & 49.4 & 87.8 & 87.5 & 84.2 \\ 
        Naive + EGAN & 85.6 & 79.8 & 0.80 & 91.5 & 46.6 & 88.6 & 88.8 & 83.7 \\ 
        Naive + Ensemble & 86.0 & 81.1 & 0.81 & 90.8 & 53.0 & 88.4 & \textbf{89.2} & 84.1 \\ 
        SleepEGAN & \textbf{86.8} & \textbf{81.9} & \textbf{0.82} & \textbf{91.7} & \textbf{53.6} & \textbf{89.2} & 89.1 & \textbf{86.1} \\ \hline
    \end{tabular}
    }
\end{table}

Table \ref{ta3} shows that the prediction performances of SleepEGAN without using EGAN and/or ensemble need to be further improved, especially for the F1-Score of N1. The Naive method has employed the weighted cross-entropy loss function as well as data augmentation  without using EGAN. Obviously, these simple strategies cannot significantly improve the prediction performance for the N1 stage. Then, we use EGAN to generate fake EEG samples in N1, but the performance is still inferior to that of SleepEGAN. Although the distribution of the training data is balanced in this scenario, the optimal model parameter selected by the validation set may not have good generalization ability in the test set due to individual heterogeneity. Therefore, we add a cost-free ensemble learning step, resulting in SleepEGAN, which improves not only the prediction accuracy of N1 but also the overall accuracy. 

\subsection{Sensitivity analysis for the number of classifiers in the ensemble learning}\label{S3-6}

The number of base learners $M$ is a hyper-parameter in the ensemble procedure and needs to be specified beforehand. We expect this hyper-parameter to be overly insensitive with respect to prediction performance. Therefore, we choose the dataset Sleep-EDF-20 for parameter sensitivity experiments. We fix the other parameters and vary $M \in \{5,6,7,8,9,10\}$ to investigate the fluctuation of its prediction result shown in Table \ref{ta4}.

\begin{table}[!ht]
    \centering
    \caption{Sensitivity analysis conducted on Sleep-EDF-20 dataset.}\label{ta4}
    \resizebox*{110mm}{!}{
    \begin{tabular}{ccccccccc}
    \hline
        \multirow{2}{*}{\# Classifiers} & \multicolumn{3}{c}
          {Overall Metrics} & \multicolumn{5}{c} {Per-class F1-Score (F1)} \\ \cline{2-9}
        & Acc & MF1 & $\kappa$ & W & N1 & N2 & N3 & REM \\ \hline
    M=5 & 86.5 & 81.5 & 0.82 & 91.4 & 51.9 & 89.1 & 89.4 & 85.6 \\ 
    M=6 & 86.7 & 81.7 & 0.82 & 91.6 & 52.8 & 89.1 & 89.3 & 85.6 \\ 
    M=7 & 86.9 & 81.9 & 0.82 & 91.6 & 53.0 & 89.4 & 89.5 & 86.2 \\ 
    M=8 & 86.7 & 81.8 & 0.82 & 91.6 & 53.0 & 89.2 & 89.3 & 85.8 \\ 
    M=9 & 86.9 & 81.9 & 0.82 & 91.7 & 53.2 & 89.3 & 89.2 & 86.3 \\ 
    M=10 & 86.8 & 81.9 & 0.82 & 91.7 & 53.6 & 89.2 & 89.1 & 86.1 \\ \hline
    \end{tabular}
    }
\end{table}

We observe that the parameter $M$ hardly affects the prediction performance of our model. For all values of $M$, the overall accuracy does not vary by more than 0.4\%. Thus, our model is very robust to the hyper-parameter $M$, and the user can choose it arbitrarily in light of the experiment's purpose.

\section{Conclusion}\label{S4}

We propose a new GAN-enhanced ensemble deep learning model, called SleepEGAN, for sleep stage classification with imbalanced classes and individual
heterogeneity from raw single-channel EEG signals. The proposed SleepEGAN outperforms several existing deep models for sleep stage classification on three popular sleep datasets. The success of SleepEGAN is mainly attributed to two aspects: first, we employ EGAN to generate fake EEG samples for the minority class so that the data become balanced during the training process; second, we develop a cost-free ensemble algorithm to reduce the estimation variance caused by individual heterogeneity, and hence it enhances the robustness of our model. Through ablation experiments, we find that these two strategies are effective to improve classification performance. Finally, we perform a sensitivity analysis on the number of base learners in the procedure of ensemble learning and show that our model works reasonably well using an arbitrary hyperparameter in a given range.

In addition, it is noteworthy that the fake EEG signals generated by our EGAN are quite similar to the real EEG signals. The EGAN method can successfully learn the temporal and transitional structure of the EEG signals, and it has potential applications in signal recognition \cite{stallkamp2012man}, signal processing \cite{hayes1996statistical,sanei2013eeg}, signal synthesis \cite{smith2000phytochromes}, among others. These can be interesting future research topics to explore. 

\section*{Declaration of Competing Interest}  \label{S7}
The authors declare that they have no known competing financial interests or personal relationships that could have appeared to influence the work reported in this paper.

\section*{Acknowledgements}
We acknowledge the support in part by the U.S. NSF grants DMS-17-12558, DMS-20-14221, Postgraduate Scientific Research Innovation Project of Hunan Province CX20200148 and National Statistical Science Research Selection (General) Project 2021LY042.

\bibliographystyle{elsarticle-num}
\bibliography{Bibliography}

\end{document}